\documentclass[prl,aps,twocolumn,showpacs]{revtex4}

\usepackage{color,amsmath,graphicx,epsfig}

\begin{document}

\def\pd#1#2{\frac{\partial #1}{\partial #2}}
\def\bk{{\bf k}}
\def\bx{{\bf x}}

\title{\bf Dissipative dynamics of superfluid vortices at non-zero temperatures}
\author {Natalia G. Berloff and Anthony J. Youd}
\affiliation {Department of Applied Mathematics and Theoretical Physics,
University of Cambridge,  Cambridge, CB3 0WA\\
}
%\date{14 March 2007}

\begin {abstract}
We consider the evolution and  dissipation of vortex rings in a condensate at
non-zero temperature,  in  the context of the classical field
approximation, based on the defocusing nonlinear
Schr\"odinger equation.  The temperature in such a system is fully
determined by the total number density and the number density of the
condensate.  A vortex ring is introduced into a condensate in a state of
thermal equilibrium, and interacts with non-condensed particles. These
interactions lead to a gradual decrease in the vortex line
density, until the vortex ring completely disappears. We show
that the square of the  vortex line length changes linearly with time, and
obtain the corresponding universal decay law. We relate this to mutual
friction coefficients in the fundamental  equation of vortex motion in superfluids.

\end{abstract}
\maketitle

The processes of self-organization, formation of large-scale coherent localized
structures and  interactions of these structures with small-scale fluctuations
are at the heart of nonlinear sciences, ranging from classical
turbulence, superfluids, ultracold gases and Bose--Einstein
Condensates (BECs), to the formation of the early Universe.
Turbulence is characterised by the co-existence of motions with many
length and time scales described by many degrees of freedom. The key to our
understanding of turbulence is to elucidate  physics of interactions between
large scales (eg. large eddies)  and small scales (eg. turbulent
fluctuations), and to develop mathematical models that account for the effects
of small scales without actually solving for them.  According to the
Landau description, superfluid $^4$He consists of the ground state and the
excitations --- quasiparticles drifting on top of the ground state. In
the language of relativistic quantum fields this corresponds to vacuum and
matter (eg. gravity waves interacting with a vacuum). Indeed, there
is a close relationship between superfluid hydrodynamics and quantum gravity,
so that at some level of hierarchy of parameters the interactions of the
quantum vacuum and matter can be described  by the defocusing nonlinear
Schr\"odinger (NLS) equation ~\cite{vol-06}. The dynamics of Bose condensates
depends on the energy exchange between the condensed and non-condensed parts of
the gas. Again, the NLS equation (reformulated as the Gross--Pitaevskii
(GP) equation~\cite{GP}) describes equilibrium and dynamical properties of BEC
as well as the formation of BEC from a strongly degenerate gas of
weakly interacting bosons \cite{ly,ks2}. The formation of the large-scale
coherent localized ground state (condensate) from a non-equilibrium initial
state has been studied in a number of papers addressing different stages
of the formation.  Weak turbulence theory has been used to predict the
self-similar evolution of the field in the regime of random phases of Fourier
amplitudes~\cite{zah,svist}, the transition from  the regime  of weak
turbulence to superfluid turbulence  via states  of strong turbulence in the
long-wavelength  region of  energy space~\cite{bs}, and the final
stage, resulting in the formation of a genuine
condensate~\cite{burnett,josserand-05}. The related question about the effect
of finite temperature on the BEC dynamics  has also been addressed
recently~\cite{gajda-07}. For instance, it was shown that the presence of the
thermal cloud in a trapped condensate creates an effective dissipation that
forces a single vortex  to  move away from the centre and
disappear~\cite{gajda-03}. 

The problem of the vortex tangle interacting with the normal fluid (thermal
cloud) is the key question in  superfluid turbulence. The Landau two-fluid
theory of superfluidity pre-dated the discovery of quantised vortex lines and
therefore omitted significant dynamical effects. This was remedied--- in
the limit in which the mean spacing between the vortex lines is small compared
with any other length scale  of interest --- by HVBK theory
\cite{hvbk1,hills}.  In this limit, the superfluid vorticity is treated as a
continuum, but the discrete nature of the vorticity gives rise to an extra
force  on the superfluid component, arising from the tension in the vortex
lines. This term is absent from the classical Euler equation of
motion for an inviscid fluid.  The vortex lines also create a force  of mutual
friction between superfluid and normal fluid in addition to the mutual
friction included by Landau in his equations, and represents the effects of
collisions of the quasiparticles with the vortex cores. Such forces were
introduced into the Landau model in an {\it ad hoc} way. This Letter is the
first attempt to study the effect of these collisions quantitatively: we shall
find {\it the vortex line decay law  at non-zero temperature in the context of
the defocusing NLS equation}. The NLS equation is a good starting point,
as the non-dissipative Landau two-fluid model can be obtained from the
equations of conservation of mass and  momentum for a one-component barotropic
fluid using a general expression for the internal energy functional of the
density~\cite{puttRob-83}. 
Through the
Madelung transformation the NLS equation can be written  in that form.
Analogously, the transport coefficients in the Landau model  have been obtained
directly from the NLS equation by following the Chapman--Engskog
expansion~\cite{kirk-85}. Note that the separation of scales needed to carry
out the derivation of the Landau two-fluid model from the NLS equation does not
allow the inclusion of vortices as part of the ground state. It is natural,
therefore,  to attempt to derive the corresponding effects of the interactions
of vortices with the quasiparticles directly from the NLS equation.
 
We consider the normalised defocusing NLS equation for the complex
function $\psi$ \cite{GP}:
\begin{equation}
i \partial_t \psi \, = \, -\nabla^2 \psi \, + \, |\psi|^2 \psi.
\label{NLS}
\end{equation} 
\noindent
The dynamics conserves the total number of particles $N= \int  |\psi|^2 d{\bf
x} $, and the total energy $E = \int \left( |\nabla\psi |^2   +  \frac{1}{2}
|\psi|^{4 }\right) d{\bf x} $. We consider the uniform discrete system of
volume $V={\cal N}^3$, which is  a periodic box on a computational grid with $128^3$
discrete points.

Our goal is to determine the universal decay law for the vortex line density in
the entire range of temperatures from 0 to the critical temperature of
condensation, $T_\lambda$. Our approach consists of three essential
steps. We aimed to: (1) achieve the thermal equilibrium state
for the given number of particles and given energy, starting from a
non-equilibrium stochastic initial condition for the wavefunction $\psi$;
(2) introduce a vortex ring into this state and follow its decay via
interactions with non-condensed quasiparticles; (3) relate the decay rate to
the temperature at equilibrium, where we derive the expression for the
relative temperature, $T/T_\lambda$, as a function of the total number density,
$\rho=N/V$, and the number density of the condensate, $\rho_0$.

We performed large scale numerical simulations of  Eq. (\ref{NLS}) starting
from a strongly non-equilibrium initial condition\cite{bs}, where the phases
of the complex  Fourier amplitudes $a_{\bf k}(t)
 =\int \psi({\bf x},t)e^{-i{\bf p}\cdot {\bf x}}\, d{\bf x}$   
are distributed randomly at $t=0$. Here the momentum ${\bf p}$ takes
quantised values ${\bf p}=(2\pi/{\cal N}){\bf n}$ with ${\bf
  n}=(0,0,0), (\pm 1,0,0), \cdot\cdot\cdot$.

This initial state describes a  weakly
interacting Bose gas that is so  rapidly cooled below the critical BEC
temperature that the particles remain in a strongly non-equilibrium
state. The kinetics of the initial weak turbulent state  has been analysed in
\cite{zah,svist,kagan} discovering a quasiparticle cascade from high energies
to low energies in the wave number space.  The ordering of the system and the
violation of the assumptions of  weak turbulence  occurs very rapidly in a
low-energy part of the spectrum, with the formation of a
quasi-condensate consisting of a tangle of quantised vortices.  The vortex
tangle decays as the system reaches a state of thermal equilibrium with
some portion ($\rho_0\equiv|a_{\bf 0}|^2/V$) of particles occupying the
zero momentum state (genuine condensate) and the rest of the
non-condensed particles being distributed according to  the
Rayleigh--Jeans equilibrium distribution~\cite{ZakhBook}, modified
by the presence of nonlinear interactions with the
condensate~\cite{josserand-05}: 
\begin{equation}
  |a^{eq}_{{\bf p}\ne {\bf 0}}|^2= \frac{T}{\omega_B(p)},
\label{equil}
\end{equation}
where $T$ is the temperature and $\omega_B(p)$ is the Bogoliubov
dispersion relation (see below). An
ultraviolet cut-off for this distribution appears naturally through the spatial
discretization of the NLS equation.  The numerical scheme consists of
fourth-order finite difference discretization in space and fourth-order
Runge--Kutta in time, so it is globally fourth-order
accurate. This scheme corresponds to the Hamiltonian system in the discrete
variables $\psi_{jkn}$,  such that 
\begin{equation}
i\dot{\psi}_{jkn}=
\frac{\partial H}{\partial \psi_{jkn}^*}, \qquad j,k,n =1,...,{\cal N},
\end{equation}
where 
\begin{eqnarray}
&H&= \sum_{jkn}
  \bigl(\psi_{jkn}^*[\textstyle{\frac{1}{12}}\Psi_2-\textstyle{{\frac{4}{3}}}\Psi_1
  +\textstyle{\frac{15}{2}}\psi_{jkn}
]+\textstyle{\frac{1}{2}}|\psi_{jkn}|^4\bigr)
\label{h}
\end{eqnarray}
with $\Psi_2=\psi_{j+2,k,n}+\psi_{j-2,k,n}
+\psi_{j,k+2,n} +\psi_{j,k-2,n}+\psi_{j,k,n+2}+\psi_{j,k,n-2}$ and $\Psi_1=\psi_{j+1,k,n}
+ \psi_{j-1,k,n}+\psi_{j,k+1,n}+\psi_{j,k-1,n}
+\psi_{j,k,n+1}+\psi_{j,k,n-1}$. 

The thermodynamic description of the condensation process has been obtained in
\cite{josserand-05} by adapting the Bogoliubov theory of a weakly
interacting Bose gas~\cite{bog} to the classical system (\ref{NLS}). We follow
the same basic idea to derive expressions for the energy and
non-condensed part of the discretised energy (\ref{h}) written in terms
of the Fourier amplitudes $a_{\bf p}$ as
\begin{equation}
H = \sum_{\bf p} K_2(p)  a^*_{\bf p} a_{\bf p} +
\frac{1}{2V}
\hspace{-0.5cm}\sum_{{\bf p}_1 , {\bf p}_2 , {\bf p}_3 , {\bf p}_4} \hspace{-0.5cm} a^*_{{\bf
p}_1} a^*_{{\bf p}_2} a_{{\bf p}_3} a_{{\bf p}_4} \delta_{{\bf p_1}\!
+\! {\bf p_2}\! -\! {\bf p_3}\! -\!  {\bf p_4}},\label{hamiltonian}
\end{equation}
where $\delta_{\bf p}$ is the Kronecker delta symbol and
\begin{equation}
  K_2(p)=\frac{2}{3}\sum_{i=1}^3\sin^{2}(p_i/2)(7-\cos(p_i)).
\label{k2}
\end{equation}

The Bogoliubov transformation $b_{\bf p} = u_{\bf p} a_{\bf p} - v_{\bf
p} a^*_{-{\bf p}}$, such that $u_{\bf p} = 1/\sqrt {1-Q_{\bf p}^2}$ and $v_{\bf
p} = Q_{\bf p}/\sqrt {1-Q_{\bf p}^2}$ with $Q_{\bf p} =  [- K_2 - 2 \rho_0 +
\omega_B(p)]/ \rho_0$ diagonalises the term in \eqref{hamiltonian}, which
is quadratic in $a_0$, to ${\sum_{\bf p}}' \omega_B(p)\, b_{\bf p}^* b_{\bf p} $,
where $ \sum_{\bf p}'$ excludes the ${\bf p}={\bf 0}$ mode. Here $\omega_B(p) =
\sqrt{ K_2^2+ 2 \rho_0  K_2}$ is the  Bogoliubov-type dispersion relation.

Using the equilibrium distribution of the uncondensed particles (\ref{equil})
the non-condensed number density can  then be expressed in terms of the
basis used in this diagonalisation as
\begin{equation}
\rho-\rho_0  
= \frac{T}{V} {\sum_{\bf p}}' \frac{K_2(p) + \rho_0}{\omega_B^2(p)}.
\label{n0bog}
\end{equation}
The discretised energy density $H/V$ in the new basis takes the form
\begin{equation}
\frac{H}{V} =  
\frac{1}{2}\left[ \rho^2+ (\rho-\rho_0)^2 \right]
+  \frac{T}{V}  {\sum_{\bf p}}'  1.
\label{tbog}
\end{equation}
The Eqs.~\eqref{n0bog}--\eqref{tbog} are analogous to Eqs.~(8)--(9)
of \cite{josserand-05} but modified for the discrete Hamiltonian discretization
(\ref{h}).  Given the energy density, $H/V$, and the total number density,
$\rho$,  one can  determine the temperature, $T$, at equilibrium and the number
density of the condensed particles, $\rho_0$, from Eqs. (\ref{n0bog}) and
(\ref{tbog}). The condensate fraction $\rho_0/\rho$ as a function of the
energy density $H/V$ is shown in FIG.1. This figure can be compared with
FIG.2 of~\cite{josserand-05} for the spectral representation of the total
energy. The analytical formulae
(\ref{n0bog})--(\ref{tbog}) predict the subcritical behaviour of
condensation, whereas the numerics does not support this
conclusion, as shown in the insert of FIG.1. We use a linear
approximation for small $\rho_0$ to determine the critical maximum energy for
condensation as shown in the insert. This energy is then used to
determine the  critical temperature for condensation $T_\lambda$ ($=T$
for $\min H/V$ for which $\rho_0=0$) from
(\ref{n0bog})--(\ref{tbog}). We found  a  phenomenological formula that
determines $T/T_\lambda$ as a function of $\rho_0$ and $\rho$ as
\begin{equation}
\frac{T}{T_\lambda}=1 - \Bigr(1 -
\alpha\sqrt{\rho}\Bigl)\frac{\rho_0}{\rho}-\alpha\sqrt{\rho}\Bigr(\frac{\rho_0}{\rho}\Bigl)^2,
\label{tt}
\end{equation}
where $\alpha$ is the only fitting parameter that we found as $\alpha =
0.227538$.  The  insert in FIG.1 shows the graph of $T/T_\lambda$ as a
function of $\rho_0/\rho$ for $\rho=1/2$. Eq. (\ref{tt}) gives an excellent fit
to the values computed from (\ref{n0bog})--(\ref{tbog}) across all the
values of $\rho_0$ and $\rho$.

In order to analyze the decay of the vortex line length at non-zero
temperatures, we insert a vortex ring into a state of  thermal equilibrium and
follow its decay due to the interactions with the non-condensed particles. The
condensate healing length, which determines the size of the vortex core, is
calculated based on the density of the condensate, and in our
non-dimensional units is $\xi=1/\sqrt{\rho_0}$. In healing lengths, the
radius of the ring is set to $R_0=10$. 
The new initial state is
$\psi_v(t=0)=\psi_{\rm eq}* \psi_{\rm vortex}$, where $\psi_{\rm eq}$ is the
equilibrium state and $\psi_{\rm vortex}$ is a wavefunction of the vortex
ring~\cite{pade}. The vortex line length, $L$, is calculated as a function of
time with high frequencies being filtered out from the field $\psi$, according
to $\tilde a_{\bf p}=a_{\bf p}*\max(\sqrt{1-p^2/p_c^2},0)$, where the cut-off
wavenumber is chosen as $p_c=10 (2\pi/{\cal N})$ ~\cite{kk}. The first important conclusion of
our numerical simulations is that {\it at all temperatures, the square of
the vortex line length decays linearly with time,}
\begin{equation}
\frac{dL^2}{dt}=-\gamma(\rho, T/T_\lambda),
\label{gamma}
\end{equation}
where $\gamma$ does not depend on $t$. FIG. \ref{ring} shows this dependence
for various temperatures. The actual isosurfaces of the decaying vortex line
are shown in the inserts.

\begin{figure}[t!]
\caption{(colour online) Condensate fraction, $\rho_0/\rho$, as a function of
  the  energy density as obtained from the numerical simulations (points) and
  from the analytical expressions (\ref{n0bog})--(\ref{tbog}) (solid
  line). The inserts show (a) the plot of  $T/T_\lambda$ as a function of
  $\rho_0/\rho$, obtained using Eqs.
  (\ref{n0bog})--(\ref{tbog}) or Eq. (\ref{tt})  and
  (b) subcritical condensation predicted by Eqs.
  (\ref{n0bog})--(\ref{tbog}) (black  line), the linear approximation
  used to obtain the critical temperature of condensation (gray (red)
  line), and numerical calculations (blue dots). The total
  number density is $\rho=1/2$. }
\centering
\bigskip
\epsfig{figure=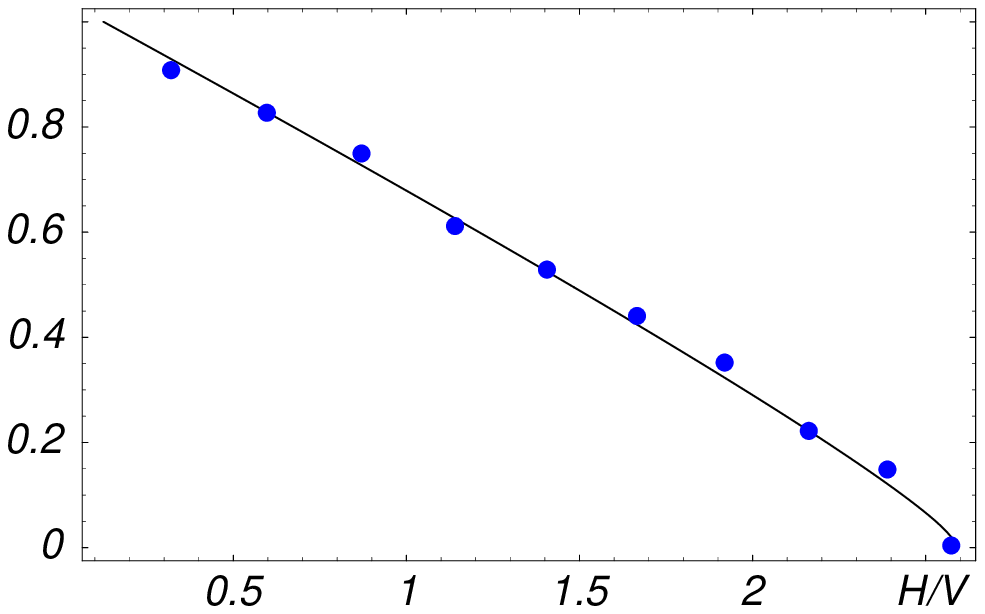, height = 2.2 in}
\begin{picture}(0,0)
\put(-115,160)  {\Large $\frac{\rho_0}{\rho}$}
\put(20,100) {\epsfig{figure=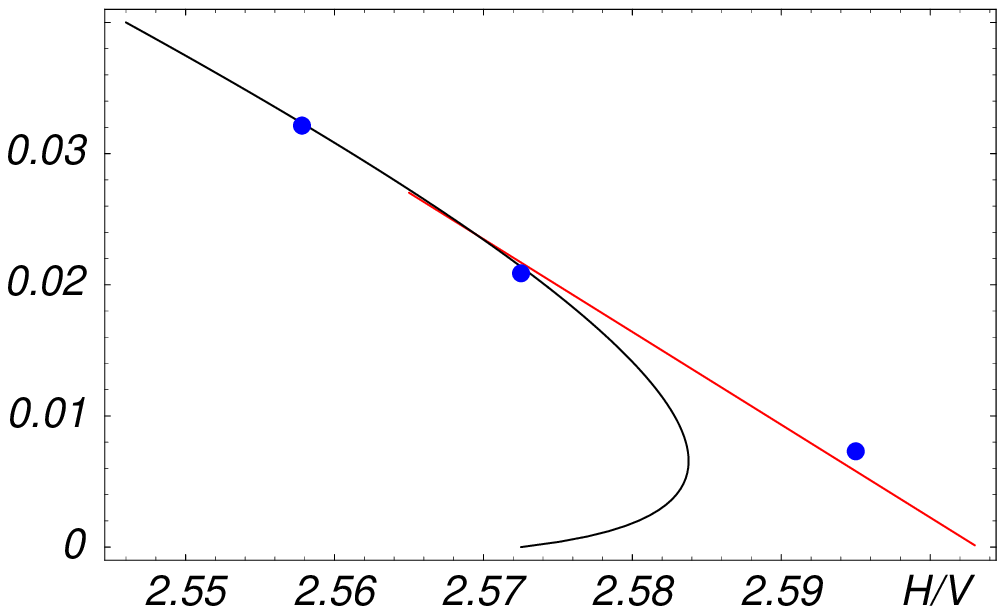, height = 0.8 in}}
\put(40,120) {(b)}
\put(-80,40) {\epsfig{figure=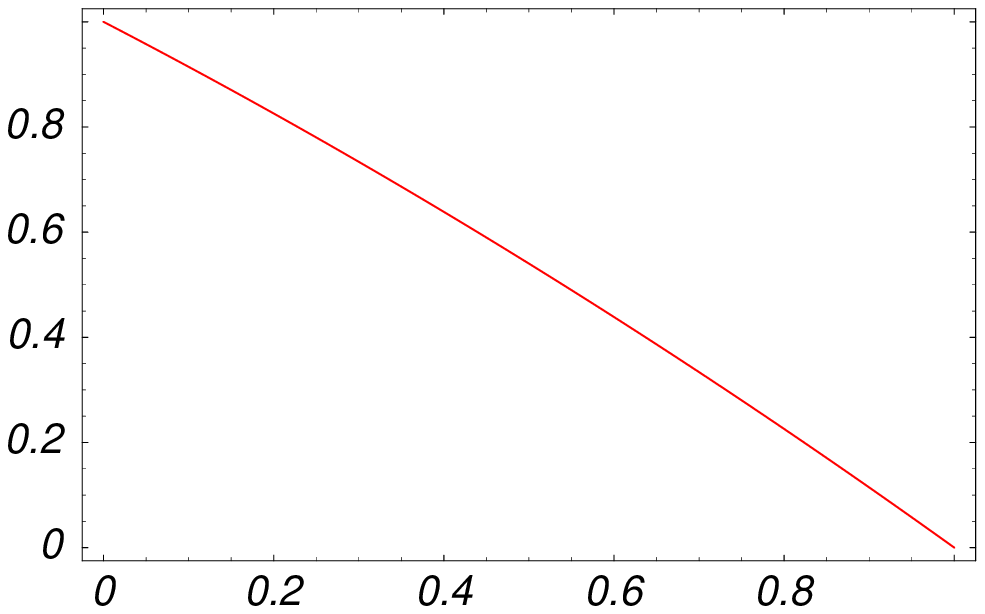, height = 0.8 in}}
\put(-60,60) {(a)}
\put(20,155) {\small $\frac{\rho_0}{\rho}$}
\put(0,40) {\small $\rho_0/\rho$}
\put(-85,100) {\small $\frac{T}{T_\lambda}$}
\end{picture}
\label{Erho}
\end{figure}

\begin{figure}[t!]
\caption{(colour online) The decay of the square of the vortex line length as
  a function of time for various $T$ indicated next to the graphs. The fit to
  the linear function is shown by the gray (red) lines. The inserts show
  isosurface plots of the vortex line (for filtered fields $\psi$; see text)
  for $T=0.52 T_\lambda$ at time=130 (left) and time=1300 (right); between
  these two times the vortex line length is reduced by a factor of
  2. The perturbations to the vortex line due to collisions with
  non-condensed particles are clearly seen on the left insert. These
  collisions generate Kelvin waves that also radiate energy to sound. The total
  number density is $\rho=1/2$. }
\centering
\bigskip
\epsfig{figure=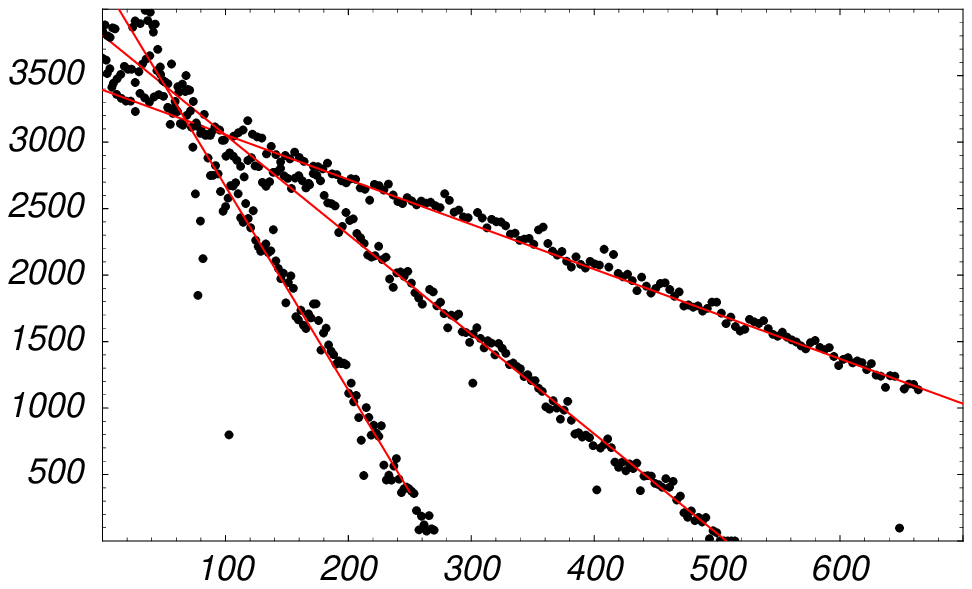, height = 2. in}
\begin{picture}(0,0)(100,100)
\put(-140,230)  { \Large $L^2$}
\put(-15,185) {\epsfig{figure=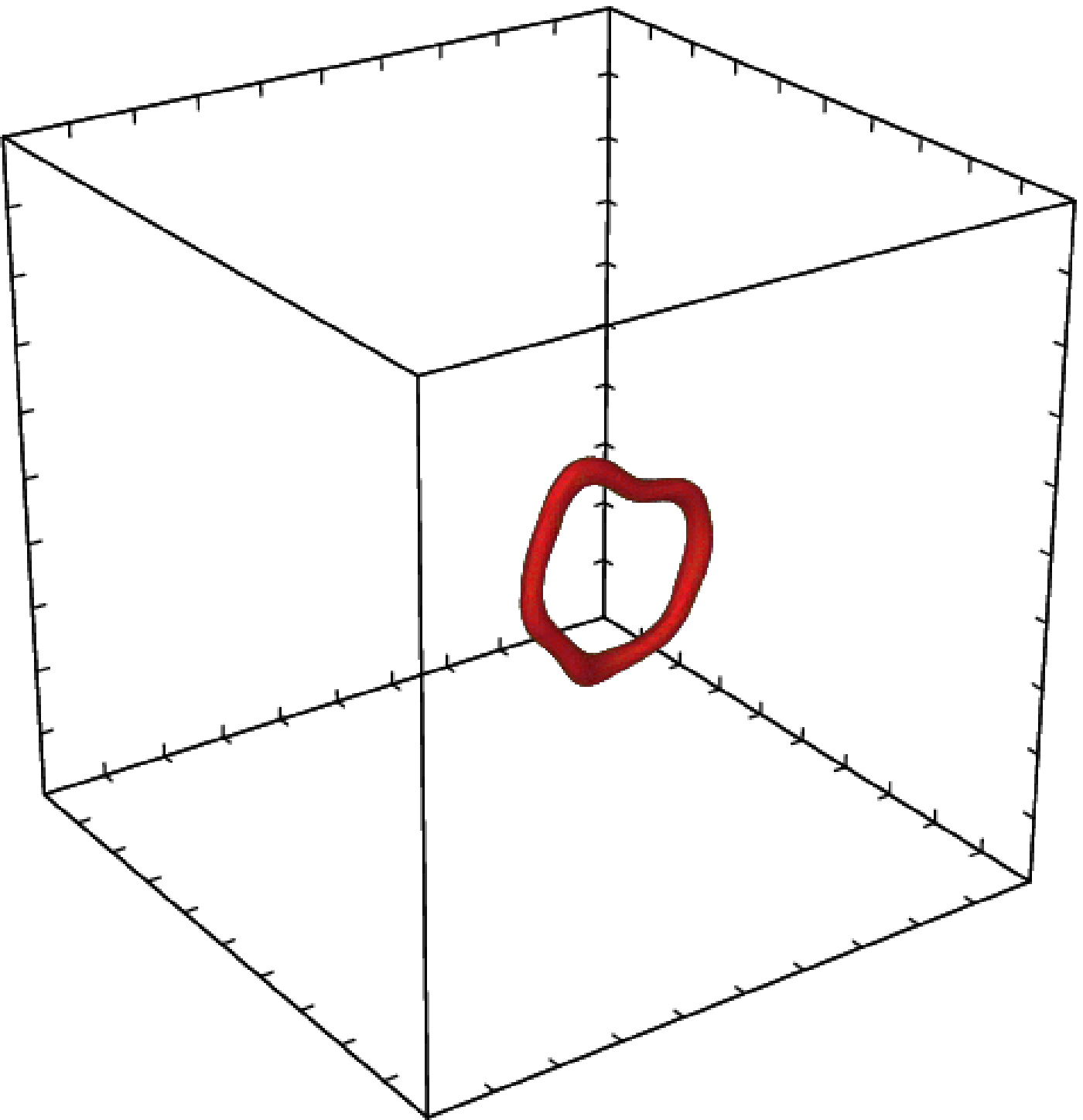, height = 0.7 in, clip=}}
\put(35,185) {\epsfig{figure=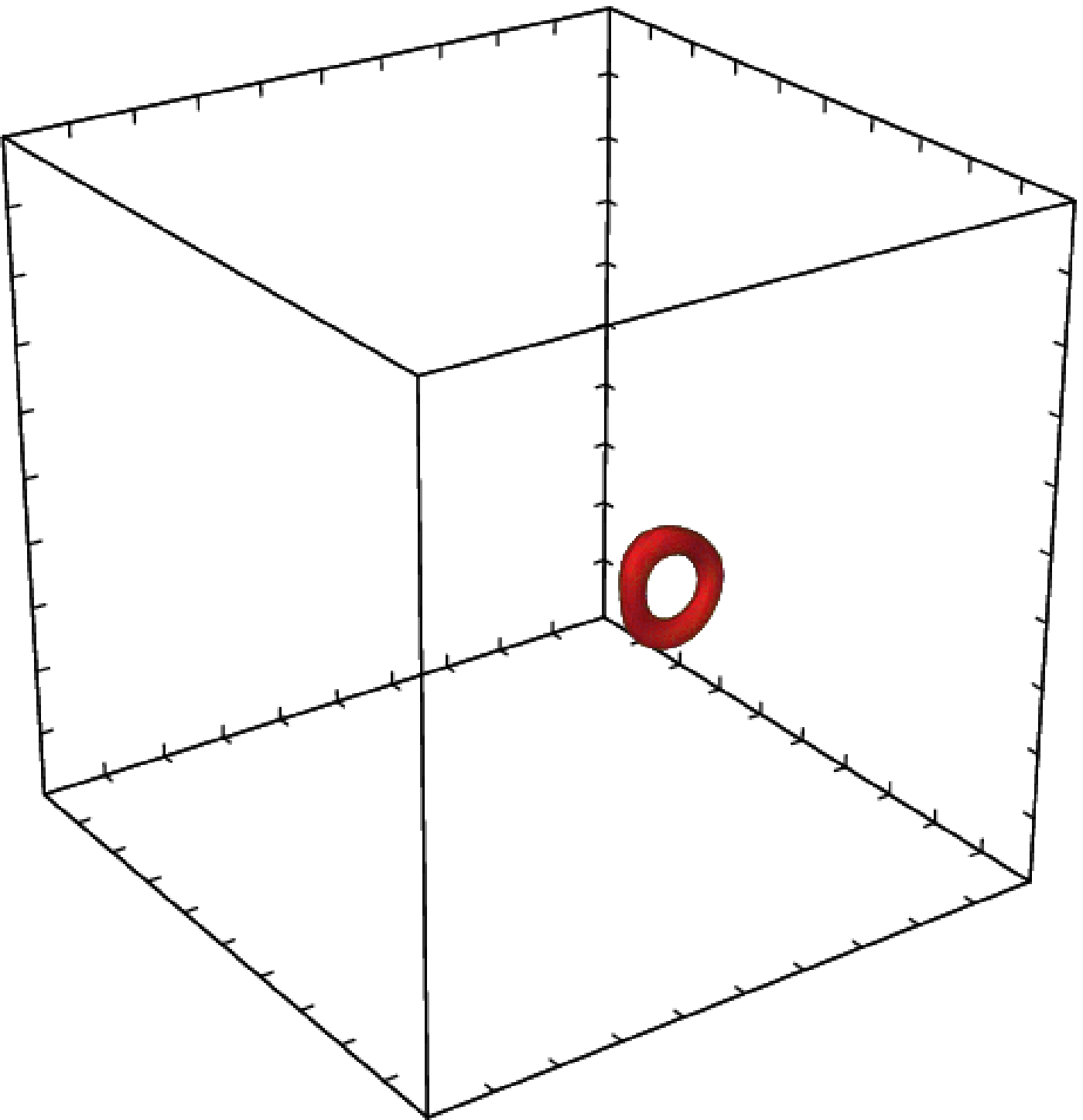, height = 0.7 in}}
\put(20,130) {\small $0.44T_\lambda$}
\put(45,145) {\small $0.27T_\lambda$}
\put(-80,130) {\small $0.63T_\lambda$}
\put(70,100) {\Large time}
\end{picture}
\label{ring}
\end{figure}
This result agrees with  predictions of the  HVBK theory for
superfluid helium \cite{hvbk1}
according to which the fundamental equation of the motion of a vortex
line, ${\bf v_L}$,  is given by (see also page 90,
Eq. (3.17) of \cite{donnelly})
\begin{equation}
{\bf v_L}={\bf v_{sl}}+ \alpha{\bf s'}\times({\bf
  v_n-v_{sl}})-\alpha'{\bf s'}\times[{\bf s'}\times({\bf v_n-v_{sl}})],
\label{vl}
\end{equation}
where ${\bf v_{sl}}$  is the local superfluid velocity that consists
of the ambient superfluid flow velocity and the self-induced vortex
velocity ${\bf u_i}$, ${\bf v_n}$ is
the normal fluid velocity, ${\bf s}$ is a position vector of
a point on the vortex and ${\bf s'}$ is the unit tangent at that
point. Mutual friction parameters $\alpha$ and $\alpha'$ are {\it ad
  hoc} coefficients in the HVBK theory that are functions of $\rho_n,
\rho,$ and $T$ only. Eq.~(\ref{vl}) is a general and universal equation used to
follow the evolution of three-dimensional vortex motion in an
arbitrary flow. When formulated for a single vortex ring Eq.~{\ref{vl}
  reads $d R/dt=-\alpha u_i$, where  $u_i=\kappa[\log(8R/\xi) - \delta
  + 1]/(4 \pi R)$ and $\delta$ is the vortex core parameter. For the GP
  vortices $\delta\approx 0.38$ \cite{GP}. In  dimensionless units used in our paper 
$
u_i=[\log(8R) - \delta
  + 1]/R.
$
After integration of the equation for ${\dot {R}}$ we get 
$\alpha t=(R_0^2-R^2)/[2(\log(8 \widehat R)+\delta -1)],
$
where $\widehat R$ is the mean radius of the ring.
When this  is compared with (\ref{gamma}) we get the following
relationship between $\gamma$ and $\alpha$:
$\gamma=8\pi^2(\log(8 \widehat R)+\delta-1) \alpha.$
From our numerics we obtained a general result valid across all ranges
of temperatures and total densities:
$
\gamma \approx K \rho (T/T_\lambda)^2, 
$
where $K\approx 68$. Note that for a GP condensate $T/T_\lambda \approx
\rho_n/\rho$ to the first order (see insert (a)  of FIG.1), so
alternatively, we can write
$\gamma \approx K_1 \rho_n (T/T_\lambda)$
FIG.~\ref{gammafig} shows the comparison of the numerically calculated
$\gamma/\rho$  and the  quadratic fit $K (T/T_\lambda)^2$. Thus, we
found that the mutual friction coefficient in condensate superfluids
is given by $\alpha\approx K_2 \rho_n (T/T_\lambda)$.
\begin{figure}[t!]
\caption{(Color online) Values of $\gamma/\rho$  as a
  function of temperature $T/T_\lambda$ for various values of the total number
  density $\rho$ depicted in various shades of gray (in various
  colours): $\rho=1/2$ (dark (red)), $\rho=1/4$ (light (green)) and
  $\rho=3/4$ (medium (blue)). The plot of the quadratic fit  $\gamma/\rho=68 (T/T_\lambda)^2$ is given by the dashed line.
  The relative temperature is calculated using Eq.~(\ref{tt}). The
  result is not  sensitive to whether we use  the values of
  $\rho$ and $\rho_0$ that  correspond to the state of thermodynamical
  equilibrium before the introduction of the vortex ring or  after the vortex ring disappears and the system
  equilibrates. The insert 
shows the distance travelled by a vortex ring as a function of time for $T/T_\lambda=0.27$ (red dots -- distances
calculated using $dz/dt=u_i$, black line using numerics). Curves
  depart when the vortex ring
becomes small in radius and the analytical formula  is no longer  accurate
  approximation of the vortex velocity. }
\centering
\bigskip
\epsfig{figure=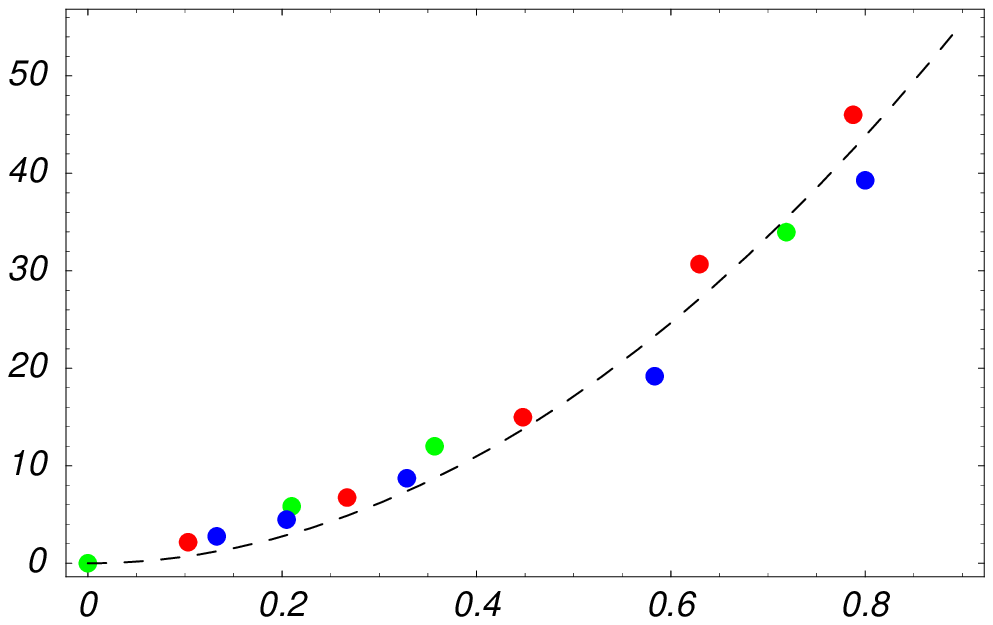, height = 2. in}
\begin{picture}(0,0)(100,100)
\put(-140,250)  { \Large $\gamma/\rho$}
\put(75,100) {\Large $T/T_\lambda$}
\put(-110,170) {\epsfig{figure=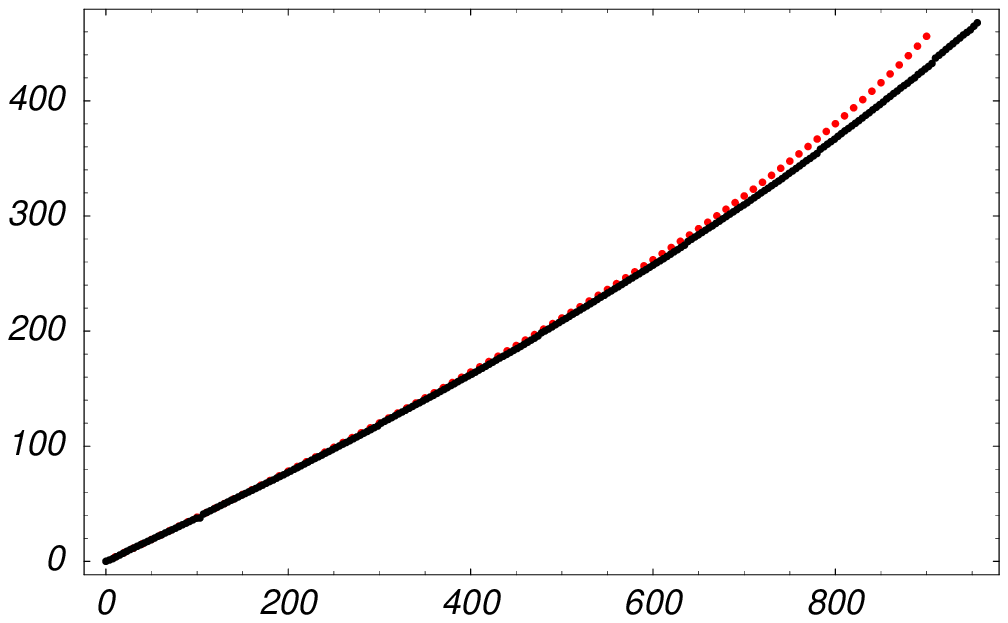, height = 0.9 in}}
\put(-10,168) {\footnotesize time}
\put(-110,235) {\footnotesize Z}
\end{picture}
\label{gammafig}
\end{figure}

The existence of the transverse force on superfluid vortices which is
parametrised by the parameter $\alpha'$ has been a subject of much
debate in mid-1990s, when  calculations of 
 the classical  Magnus force applied to superfluid vortices have been offered and argued about
 \cite{alphapr}. The criticism is based on the observation that the
 classical hydrodynamic equations are inapplicable in the vortex
 core. Whether or not the details of the non-classical  vortex
 dynamics are crucial to the existence of the transverse force is
 still an open question. The estimate of $\alpha'$ can be obtained
 from our numerical procedure as following. Eq. (\ref{vl}) 
 written for a distance travelled by a single vortex ring takes form (see
 Eq.~(3.53) on page 107 of \cite{donnelly}) $dz/dt=(1-\alpha')u_i$. We
 compared the distances travelled by a
vortex ring at various temperatures obtained numerically with the
distances travelled by a vortex ring  in the
absence of the transverse force according to the analytical formula 
$d z/d t = u_i,
$
where $u_i=u_i(R(t))$ and $R(t)$  varies with time according to
(\ref{gamma}).  The insert of FIG.\ref{gammafig}
shows these distances for $T/T_\lambda=0.27$. Our calculations fail to detect any significant
presence of the transverse force for any temperature considered: the deviation from the analytical
curve is insignificant within the accuracy of (\ref{gamma}).    We plan
to perform a more  thorough analytical and numerical study of 
transverse force from a single phonon acting on a single vortex in
context of the GP model in future.

In summary, we considered the effect of temperature on the decay of
vortex line density via interactions with non-condensed particles in the context of the defocusing NLS equation. We
obtained a simple expression for the temperature at equilibrium as a function
of the total number density and the number density of the condensed
particles. Depending on these two parameters, a
vortex ring introduced into the condensate shows different
decay rates with time. We identified this decay law as linear for the
square of the vortex line length and showed the universal
dependence of the decay rate on temperature and total density. It has
been suggested that the emission of sound by vortex reconnections and
vortex motion is the only active dissipation mechanism responsible for
the decay of superfluid turbulence. The decay of superfluid turbulence
via Kelvin wave radiation and vortex reconnections was studied in the
framework of the GP equation \cite{leadbeater2}  at near zero
temperature, via collision of two vortex rings, and confirmed
that in the  Kelvin wave cascade, where energy is transferred to much shorter
wavelengths with a cut-off below a critical wavelength, the vortex
 line density  can be described by the famous Vinen equation~\cite{vinen}
$
{d (L/V)}/{d t} = -\chi {(L/V)}^2.
$
It has also been shown~\cite{ber-04} that the presence of localized finite amplitude
sound waves greatly enhances the  dissipation of
the vortex tangle, essentially changing the decay law to exponential
decay.  This Letter complements the existing Kelvin wave cascade
scenario by considering an opposite limit when there are no
reconnections, and the decay mechanism depends only on the energy exchange
with non-condensed particles. This mechanism exceeds the energy
transfer via the Kelvin wave cascade. Finally, we related our results
about a single vortex ring to the  mutual
friction coefficients in the general equation of vortex motion.

NGB acknowledges the support from EPSRC-UK. She is also very grateful
to Professor Joe Vinen for several illuminating discussions about superfluid
turbulence and his suggestion to look into the decay of vortex rings
in the context of the NLS equation.


\begin{thebibliography}{}
\bibitem{vol-06} G. Volovik, gr-qc/0612134.
\bibitem{GP} V. L. Ginzburg and L. P. Pitaevskii, Sov. Phys. JETP {\bf 7}, 858
(1958); L. P. Pitaevskii, Sov. Phys. JETP {\bf 13}, 451 (1961);
  E. P. Gross J. Math. Phys. {\bf 4}, 195 (1963).
\bibitem{ly} E. Levich and V. Yakhot, J. Phys. A: Math. Gen. {\bf 11}, 2237
(1978).
\bibitem{ks2} Yu.~Kagan and B.V.~Svistunov, Phys. Rev. Lett. {\bf 79}, 3331
(1997).
\bibitem{zah} V.E.Zakharov, S.L. Musher, and A.M.Rubenchik,
  {\it Phys. Rep.} {\bf 129}, 285 (1985) and S. Dyachenko, A.C. Newell,
  A. Pushkarev and V.E.Zakharov, {\it Physica D} {\bf 57}, 96 (1992).

\bibitem{svist} B.V. Svistunov, {\it J. Moscow Phys. Soc.} {\bf 1}, 373
(1991); Yu. Kagan and
B.V. Svistunov, {\it Zh. Eksp. Theor. Fiz.} {\bf 105},
353 (1994) [{\it Sov. Phys. JETP} {\bf 78}, 187 (1994)].
\bibitem{bs} N. G. Berloff and B. V. Svistunov, Phys. Rev. A {\bf 66}, 013603 (2002)
\bibitem{burnett} M.J.Davis, S.A. Morgan, and K. Burnett,
  {\it Phys. Rev. Lett} {\bf 87}, 160402 (2001) and {\it Phys. Rev. A} {\bf 66},
  053618 (2002)
\bibitem{josserand-05} C. Connaughton {\it et al}, {\it Phys. Rev. Lett.} {\bf
  95}, 26901 (2005).
\bibitem{gajda-07}M.Brewczyk et al {\it J. Phys. B: At. Mol. Opt. Phys.} {\bf
    40}, R1-R37 (2007) and reference within.
\bibitem{gajda-03}H.  Schmidt et al {\it J.Opt.B: Quantum
    Simiclass. Opt. }{\bf 5} S96 (2003).
\bibitem{hvbk1} H.E. 
Hall and W. F. Vinen, 
{\it Proc.\ R. Soc.\ Lond., A}{\bf 238}, 215 (1956);
I.L. Bekharevich and I.M. Khalatnikov
{\it Soviet Phys., JETP}, {\bf 13}, 643 (1961).


\bibitem{hills} R.N. Hills,  and P.H. Roberts,  {\it Archiv.\ Rat.\ Mech.\
\& Anal.}, {\bf 66}, 43 (1977a);
{\it Int.\ J. Eng.\ Sci.}, {\bf 15}, 305 (1977b);
{\it J. Low Temp.\ Phys.}, {\bf 30}, 709 (1978a);
{\it J. Phys.\ C}{\bf 11}, 4485 (1978b).



\bibitem{puttRob-83} S.J. Putterman and P.H.Roberts, {\it Physica},
  {\bf 117A}, 369 (1983).
\bibitem{kirk-85} T.R.Kirkpatrick and J. R. Dorfman {\it J. Low
  Temp. Phys.} {\bf 58}, 301 (1985); 399 (1985).

\bibitem{kagan} Yu. Kagan, B.V. Svistunov, and G.V. Shlyapnikov, {\it Zh. Eksp. Teor. Fiz.} {\bf 101}, 528 (1992)
[{\it Sov. Phys.  JETP} {\bf 75}, 387 (1992)]; Yu. Kagan and
B.V. Svistunov, {\it Zh. Eksp. Theor. Fiz.} {\bf 105},
353 (1994) [{\it Sov. Phys. JETP} {\bf 78}, 187 (1994)].
\bibitem{ZakhBook} V. E. Zakharov, V. S. L'vov and G. Falkovich,
{\it Kolmogorov Spectra of Turbulence I} (Springer, Berlin, 1992); A. C. Newell, S. Nazarenko and L. Biven, Physica D {\bf 152}, 520 (2001).
\bibitem{bog} N. N. Bogoliubov, Journal of Physics {\bf 11}, 23
  (1947).

\bibitem{pade} N.G. Berloff {\it J.
Phys. A: Math. and Gen.}, {\bf 37}(5), 1617 (2004).

\bibitem{kk} Various choices of $p_c$ taken from the interval
  $(5,15)\times(2 \pi/{\cal N})$
  give different $L$, but similar vortex line decay rates. A more
  accurate, but more computationally intensive, way to calculate the
  vortex line length in a quasi-condensate (a condensate with vortices)
  is through a time averaging of the field that removes the
  high frequencies, see the discussion in ~\cite{bs}.

\bibitem{donnelly} R.J.Donnelly ``Quantized Vortices in Helium II'',
  Cambridge University Press, Cambridge 1991.
\bibitem{alphapr} S.V.Iordanskii, {\it Sov. Phys. JETP} {\bf 22}, 160 (1966);
  E. B. Sonin, {\it Sov. Phys. JETP} {\bf 42} 469 (1975); {\it
  Phys. Rev. B} {\bf 55},  485  (1997), Ao and Thouless, {\it
  Phys. Rev. Lett.} {\bf 70}, 2158 (1993).
\bibitem{leadbeater2}
M. Leadbeater, D.C. Samuels, C.F. Barenghi and C.S. Adams,
Phys. Rev. A {\bf 67} 015601 (2003).

\bibitem{vinen}
W.F. Vinen, Proc. R. Soc. London. Ser. A {\bf 242}, 493 (1957)

\bibitem{ber-04} N.G. Berloff  {\it   Phys. Rev A}, {\bf   69} 053601 (2004)
%-----------------------------------------------------------------

\end{thebibliography}
\end{document}